\def\papertitle{Anti-aliasing of neural distortion effects via model fine tuning}
\def\paperauthorA{Alistair Carson}
\def\paperauthorB{Alec Wright}
\def\paperauthorC{Stefan Bilbao}

\documentclass[twoside,a4paper]{article}
\usepackage{etoolbox}
\usepackage{dafx_25}
\usepackage{amsmath,amssymb,amsfonts,amsthm}
\usepackage{euscript}
\usepackage[T1]{fontenc}
\usepackage[utf8]{inputenc}
\usepackage{ifpdf}
\usepackage[english]{babel}
\usepackage{caption}
\usepackage{subfig} % or can use subcaption package
\usepackage{color}
\usepackage{booktabs}
\usepackage{multirow}
\usepackage{xcolor}
\def\SB[#1]{\textcolor{blue}{\emph{#1}}}

\input glyphtounicode
\ninept

\newcounter{numauth}
\newcounter{listcnt}
\newcommand\authcnt[1]{\ifdefined#1 \stepcounter{numauth} \fi}

\newcommand\addauth[1]{
\ifdefined#1 
\stepcounter{listcnt}
\ifnum \value{listcnt}<\value{numauth}
\appto\authorslist{, #1}
\else
\appto\authorslist{~and~#1}
\fi
\fi}
\authcnt{\paperauthorB}
\authcnt{\paperauthorC}
\authcnt{\paperauthorD}
\authcnt{\paperauthorE}
\authcnt{\paperauthorF}
\authcnt{\paperauthorG}
\authcnt{\paperauthorH}
\authcnt{\paperauthorI}
\authcnt{\paperauthorJ}
\def\authorslist{\paperauthorA}
\addauth{\paperauthorB}
\addauth{\paperauthorC}
\addauth{\paperauthorD}
\addauth{\paperauthorE}
\addauth{\paperauthorF}
\addauth{\paperauthorG}
\addauth{\paperauthorH}
\addauth{\paperauthorI}
\addauth{\paperauthorJ}
\usepackage{times}
\usepackage{siunitx}

\newif\ifpdf
\ifx\pdfoutput\relax
\else
   \ifcase\pdfoutput
      \pdffalse
   \else
      \pdftrue
   \fi
\fi

\ifpdf % compiling with pdflatex
  \usepackage[pdftex,
    pdftitle={\papertitle},
    pdfauthor={\authorslist},
    pdfsubject={Proceedings of the 28th International Conference on Digital Audio Effects (DAFx25)},
    colorlinks=false, % links are activated as color boxes instead of color text
    bookmarksnumbered, % use section numbers with bookmarks
    pdfstartview=XYZ % start with zoom=100% instead of full screen; especially useful if working with a big screen :-)
  ]{hyperref}
  \usepackage[pdftex]{graphicx}
\else % compiling with latex
  \usepackage[dvips]{epsfig,graphicx}
  \usepackage[dvips,
    pdftitle={\papertitle},
    pdfauthor={\authorslist},
    pdfsubject={Proceedings of the 28th International Conference on Digital Audio Effects (DAFx25)},
    colorlinks=false, % no color links
    bookmarksnumbered, % use section numbers with bookmarks
    pdfstartview=XYZ % start with zoom=100% instead of full screen
  ]{hyperref}
\fi
\usepackage[hypcap=true]{caption}
\title{\papertitle}

\affiliation{
\paperauthorA\,\sthanks{A. Carson is funded by the Scottish Graduate School of Arts and Humanities (SGSAH)}, \paperauthorB\: 
 and \paperauthorC}
 {\href{https://www.acoustics.ed.ac.uk/}{Acoustics and Audio Group} \\ University of Edinburgh \\ Edinburgh, UK\\
 {\tt \href{mailto:alistair.carson@ed.ac.uk}{<alistair.carson, alec.wright, s.bilbao>@ed.ac.uk}}
 }

\begin{document}
% more pdf-tex settings:
\ifpdf % used graphic file format for pdflatex
  \DeclareGraphicsExtensions{.png,.jpg,.pdf}
\else  % used graphic file format for latex
  \DeclareGraphicsExtensions{.eps}
\fi

%\makeatletter
%\pdfbookmark[0]{\@pdftitle}{title}
%\makeatother

\maketitle

\begin{abstract}
Neural networks have become ubiquitous with guitar distortion effects modelling in recent years. Despite their ability to yield perceptually convincing models, they are susceptible to frequency aliasing when driven by high frequency and high gain inputs. Nonlinear activation functions create both the desired harmonic distortion and unwanted aliasing distortion as the bandwidth of the signal is expanded beyond the Nyquist frequency. Here, we present a method for reducing aliasing in neural models via a teacher-student fine tuning approach, where the teacher is a pre-trained model with its weights frozen, and the student is a copy of this with learnable parameters. The student is fine-tuned against an aliasing-free dataset generated by passing sinusoids through the original model and removing non-harmonic components from the output spectra. Our results show that this method significantly suppresses aliasing for both long-short-term-memory networks (LSTM) and temporal convolutional networks (TCN). In the majority of our case studies, the reduction in aliasing was greater than that achieved by two times oversampling. One side-effect of the proposed method is that harmonic distortion components are also affected. This adverse effect was found to be model-dependent, with the LSTM models giving the best balance between anti-aliasing and preserving the perceived similarity to an analog reference device.
\end{abstract}

\section{Introduction}
Systems for nonlinear waveshaping, filtering and amplification are essential tools in music production and performance, in particular electric guitar playing. Devices such as vacuum tube amplifiers and transistor-based fuzz pedals have been used since the 1960s to shape the sound of the electric guitar through the introduction of harmonic distortion to the spectrum. In the digital domain, the simplest distortion effects can be implemented with clipping or saturating non-linear functions. Most digital distortion effects, however, seek to model or emulate analog devices \cite{DAFXVAchapter}. Methods for virtual analog modelling of distortion effects include circuit-based white-box methods \cite{Karjalainen06, Paiva12}, and black-box modelling including neural network approaches \cite{Eichas:2016, Wright2019RNN, Wright2020, juvela2023end}.

An inherent problem with non-linear digital audio processing is aliasing distortion, which is often perceived as unpleasant artefacts, beating or noise \cite{Lehtonen_2012_saw}. The harmonics generated by deliberate clipping of a signal often exceed the Nyquist frequency, therefore causing aliasing within the audio band. The canonical method for reducing aliasing is oversampling -- processing at rates of two or more times the audio rate (see e.g. \cite{Kahles2019}). Alternative methods have also been explored, but are often limited to a certain class or subset of functions, e.g. bandlimited interpolation applied to soft-clipping functions \cite{esqueda_blamp2015}. Parker et al. \cite{Parker2016} proposed a method based on continuous-time convolution of the distorted signal with an anti-aliasing low-pass filter, known as antiderivative anti-aliasing (ADAA). This method and variants thereof work well for memoryless nonlinear functions \cite{Parker2016, bilbao2017_adaa_memoryless, bilbao2017_spectral_flatness, zhelevnov24} and virtual analog models with one or two states \cite{Holters2019_adaa, Holters2020_adaa, Albertini2020}. However, the application of ADAA to larger state-space systems or neural networks is an open research question. 

In general, anti-aliasing in the context of neural distortion effects constitutes an interesting research problem. Vanhatalo et al. \cite{Vanhatalo2024} considered various methods: using (synthetic) oversampled training data; low-pass filter placement between layers; incorporating spectral loss functions into training; and using sparse networks through model pruning. Out of these, forced sparsity was found to be a viable option for a temporal convolutional network (TCN), but was less effective for the long-short-term-memory network (LSTM) example. Furthermore, it came at the cost of reduced model accuracy \cite{Vanhatalo2024}. K{\"o}per and Holters \cite{Koper2023} proposed an anti-aliased state-trajectory network model, and whilst in some cases a reduction in aliasing was shown, the main limitation was that the model could only be trained on synthetic data, not audio from an analog device. Our previous work \cite{Chowdhury2022, Carson2024} showed that $M$ times oversampling can be implemented in LSTMs by adjusting the feedback delay length to $M$ samples. With an appropriate design of interpolation and decimation filters \cite{carson2025resamplingfilterdesignmultirate}, this can be employed to reduce aliasing, but of course comes at the expense of $M$ times more operations per input sample.

Here, we investigate a data-driven approach to reducing the aliasing caused by neural network models of distortion effects without any modifications to the model architecture itself or oversampling. This paper is structured as follows: Sec.~\ref{sec:method} outlines the proposed methodology; Sec.~\ref{sec:experiments} describes the models used as case studies; Sec.~\ref{sec:obj_results} presents objective results and spectral analysis; Sec.~\ref{sec:mushra} contains a perceptual evaluation; and Sec.~\ref{sec:conclusion} provides concluding remarks. Open source code and audio examples are available \footnote{\href{https://a-carson.github.io/dafx25_antialiasing_neural/}{https://a-carson.github.io/dafx25\_antialiasing\_neural/}}.

\section{Methodology} \label{sec:method}

\begin{figure*}[ht]
\center
\includegraphics[width=1.0\textwidth]{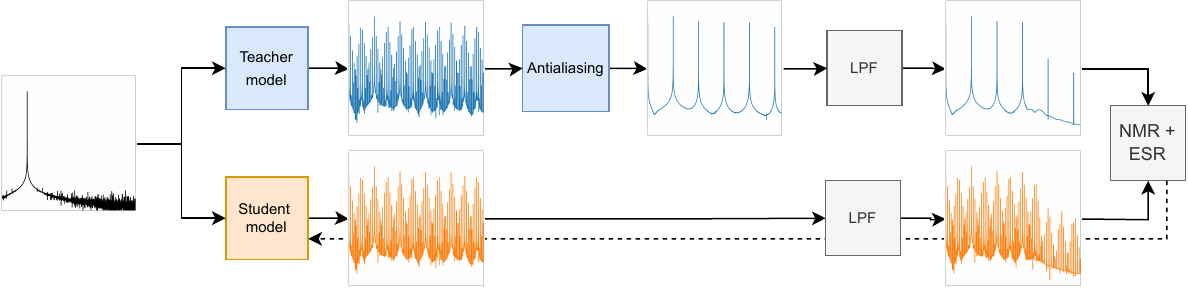}
\caption{\label{fig:fine_tune}{\it Fine-tuning procedure for anti-aliasing of the Student model. The dashed line indicates the flow of gradients to the Student parameters. Spectral plots are included for illustration only --  training operates in the time domain (except for the NMR calculation). }}
\end{figure*}

Consider an audio processing neural network of the form:
\begin{equation}\label{eq:model_basic}
    y = f(x, \theta)
\end{equation}
where $x$ is the input signal, $y$ is the output signal and $\theta$ are the model parameters or weights. Our methodology is not model-specific, so we allow $f$ to be any neural network including but not limited to recurrent neural networks (RNNs) and TCNs. Despite the fact that RNNs and TCNs can both yield perceptually convincing models of distortion effects \cite{Wright2020, Damskagg:2019:ICASSP19}, it has been shown that the non-linear activation functions, while responsible for creating the desired harmonic distortion effect, also generate aliasing \cite{Vanhatalo2024, damskagg2019real}.
% \SB[Need a reference to Wright in this sentence. Also, the model expressed above looks memoryless for a function $f$, though it's not intended to be.].
We assume that $f$ has been pre-trained against some ground truth audio output from an analog device, $y_{\rm ref}$, but that we no longer necessarily have access to the original device or the training data collected from it. For example, the model may have been obtained as open-weight from elsewhere (e.g. \cite{Wright2025}). The first step in our method is to duplicate the pre-trained model, with the original version designated as the \textit{teacher model} and the other the \textit{student model}. Formally these are defined:
\begin{subequations}
\begin{align}
    y_{\rm teach} &= f(x, \theta_{\rm teach}) \\
    y_{\rm stud} &= f(x, \theta_{\rm stud}),
\end{align}
\end{subequations}
where initially $\theta_{\rm stud} = \theta_{\rm teach} = \theta$. The weights of the teacher model are then frozen and randomised sine tones are passed through the model to produce a synthetic dataset of signals which can be decomposed into harmonics plus noise (including but not limited to aliasing noise). We assume that the ``perfect'' output signal in virtual analog modelling is purely harmonic, so remove all non-harmonic components from the teacher model output using spectral analysis and Fourier synthesis. These aliasing-free signals are then used as a target during the training of the student model weights. This ``fine-tuning" procedure is illustrated in Fig. \ref{fig:fine_tune} and details are covered in the remainder of this section. 

\subsection{Input sine tones}
The input to both the teacher and student model during training is a batch of pseudo-randomly generated sine tones, defined as:
\begin{equation}
    x[n] = A \sin(2\pi f_0 n/F_s + \phi),
\end{equation}
where $F_s$ is the sampling rate (and set to \SI{44.1}{\kHz} unless otherwise specified). The amplitude $A$ and phase $\phi$ are sampled from uniform distributions:
\begin{eqnarray}
    A \sim \mathcal{U}(0, 1), \quad \phi \sim \mathcal{U}(0, 2\pi)
\end{eqnarray}
% \SB[Do you set this amplitude range because your the network is already trained to accept normalised audio signals as inputs? Otherwise, how do you know the appropriate range?]
and the frequency is set according to:
\begin{equation}
    f_0 = 440 \cdot 2 ^ {(l - 69)/12}, \quad l \sim \mathcal{U}(21, 127).
\end{equation}
The bounds on $l$ were chosen as the range of midi notes on a standard keyboard, but here $l$ may be integer or non-integer valued.
% \SB[Is $m$ an integer here? If so, need a different notation for ${\mathcal U}$]
The sinusoids are processed through the teacher model to generate the signal $y_{\rm teach}$ which contains a mixture of harmonic distortion and aliasing noise components. 

\subsection{Aliasing removal}
The teacher model output is then post-processed to remove aliasing. First the initial samples are truncated, removing transients to leave a one-second segment. The signal is then multiplied by a Chebyshev window $w[n]$ with \SI{120}{\dB} side-lobe attenuation and the FFT taken to obtain the spectrum:
\begin{equation}
    Y_{\rm teach}[k] = \sum_{n=0}^{N-1}w[n] y_{\rm teach}[n] e^{-j2\pi k / N}.
\end{equation}
Given a sinusoidal input, the harmonic frequencies of the model output occur at, for integer $m$:
\begin{equation}
    f_m = (m+1) f_0
\end{equation}
where $0 \leq m \leq M-1$, and $M = \lfloor F_s/(2f_0)\rfloor$. The closest FFT bins to the harmonic frequencies are therefore:
\begin{equation}
    b_m = {\rm round}(f_m N / F_s)
\end{equation}
and the difference between these bins to the ``true'' bin indices are:
\begin{equation}
    d_m = f_m N / F_s - b_m.
\end{equation}
Now the complex amplitudes of the harmonic components can be extracted from the spectrum and adjusted for the scalloping loss that arises when the harmonic is an off-bin frequency \cite{Harris78_windows, zhelevnov_thesis}:
\begin{equation}
    C_m = \frac{Y_{\rm teach}[b_m]}{\sum_{n=0}^{N-1}w[n] e^{j2\pi n d_m / N}}.
\end{equation}
% \SB[This risks being very inaccurate in cases where the ideal frequency is directly between bins. Quadratic interpolation of three neighbouring amplitudes does a way better job. Is this what the scalloping loss adjustment does?]
Finally, a bandlimited alias-free version of the signal is constructed using Fourier synthesis:
\begin{equation}
    \tilde{y}_{\rm teach}[n] = B + 2 \sum_{m=0}^{M-1} |C_m| \cos\left(2 \pi f_m n + \angle C_m \right)
\end{equation}
where  $B$ is the DC offset (adjusted for the windowing gain):
\begin{equation}
    B = \frac{Y_{\rm teach}[0]}{\sum_{n=0}^{N-1}w[n]}.
\end{equation}
From here on, the tilde notation denotes the bandlimited aliasing-free version of a distorted sinusoidal signal. 

\subsection{Pre-emphasis filtering}
Pre-emphasis filtering is a common technique in black-box modelling of audio effects \cite{Damskagg:2019:ICASSP19, wright2020_perceptual}. Prior to loss computation in our model, both the target  $\tilde{y}_{\rm teach}$ and student model output $y_{\rm stud}$ are filtered by a low-pass filter (LPF) with passband and stopband edges of \SI{12}{\kHz} and \SI{16}{\kHz} respectively and \SI{80}{\dB} stopband attenuation. The justification for this is that aliasing is much more noticeable when it appears below the fundamental frequency of the desired signal and human hearing depreciates considerably above \SI{16}{\kHz} \cite{Lehtonen_2012_saw}. The filter therefore encourages very high frequency errors to be ignored during training, and initial experiments showed that this improved results. A comparison against other pre-emphasis filters was considered but not included in the scope of this work. 

\subsection{Loss function}
The loss between the student model output and the alias-free teacher output is computed as the sum of the error-to-signal ratio (ESR) and the noise-to-mask ratio (NMR). The ESR is commonly used in black-box modelling of audio effects \cite{Damskagg:2019:ICASSP19, Wright2019RNN, Wright2020}. For any arbitrary signal $\hat{s}$ and a reference signal $s$, the ESR is defined as:
% \begin{equation}
%     \mathbb{E}(S, S_{\rm ref}) = \frac {\left \lVert S - S_{\rm ref} \right \rVert_{\rm F}^2}
%     {\left \lVert S_{\rm ref} \right \rVert_{\rm F}^2}.
% \end{equation}
% where $\left \lVert \cdot \right \rVert_{\rm F}$ is the Frobenius norm. 
\begin{equation}
    \mathbb{E}(\hat{s}, s) = \frac {\sum_{n=0}^{N-1}(s[n] - \hat{s}[n])^2}
    {\sum_{n=0}^{N-1}s[n]^2}.
\end{equation}

Furthermore, in this work we use the noise-to-mask ratio (NMR) in the loss function and in evaluation. The NMR is the energy ratio between non-harmonic components and the simplified masking threshold of desired harmonic components \cite{NMR_ITU-R, KabalNMR2003} and has been used in several works as a perceptually-informed aliasing measurement\cite{Lehtonen_2012_saw, Kahles2019, zhelevnov24, zhelevnov_thesis}, and as a loss function in neural water-marking \cite{Moritz2024}. The NMR is computed between STFTs of the signal and reference signal, so let us denote the STFT of signal $s$ as $S_{k, t}$ for $k= 0, \dots, K-1$, $t = 0, \dots,  T-1$ where $K$ and $T$ are the number of frequency bins and time frames respectively. For an FFT length of $N_{\rm FFT}$ this gives $K= N_{\rm FFT}/2 + 1$ bins. 

The first step in NMR calculation is to compute the STFTs of signal $\hat{s}$ and the reference $s$ to obtain $\hat{S}_{k, t}$ and $S_{k, t}$ respectively. The \textit{noise pattern} is then computed:
\begin{equation}
    N_{c, t} = \sum_{k=0}^{K-1} U_{c, k} \Big( \omega_{k} (|\hat{S}_{k, t}| - |S_{k, t}|) \Big)^2
\end{equation}
where $\omega_k$ is a filter approximation of the human inner and outer ear; and $U_{c, k}$ are elements of a $C\times K$ matrix that maps the STFT frequencies from $K$ FFT bins to $C$ critical bands (the rows are bandpass filters for each band). The \textit{masking pattern} is computed from the reference STFT as:
\begin{equation}
    M_{c, t} = \mathcal{S} \left( \sum_{k=0}^{K-1} U_{c, k} \cdot |\omega_k \cdot S_{k, t}|^2 + \mu_c \right)
\end{equation}
where $\mu_c$ is the internal ear noise for each band and $\mathcal{S}(\cdot)$ is a function that spreads energy between the critical bands to account for frequency masking -- details of which are omitted here for brevity but the reader is referred to the work of Kabal (Section 2.8) \cite{KabalNMR2003}.
% \SB[Definition above is a little hard to follow, because you still have $k$ in there...can you define ${\mathcal S}$ somehow? It must involve a sum over $k$.]
The noise-to-mask ratio is then computed:
\begin{equation}
    {\rm NMR}(\hat{S}, S) = \frac{1}{C \cdot T}\sum_{c=0}^{C-1}\sum_{t=0}^{T-1}\frac{N_{c, t}}{M_{c, t}}
\end{equation}
For this work we adapt the MATLAB implementation provided by Zheleznov \cite{zhelevnov_thesis} into a differentiable PyTorch module, available in the accompanying code. A window and FFT size of $N_{\rm FFT} = 2048$ was used with an overlap of 50\% and $C=109$ critical bands.

Given the student model output $y_{\rm stud}$, the bandlimited teacher model output $\tilde{y}_{\rm teach}$ and their respective STFTs $S_{\rm stud}$ and $\tilde{S}_{\rm teach}$, the loss function used in fine tuning is therefore:
\begin{equation}\label{eq:loss}
    \mathcal{L} = \mathbb{E}(y_{\rm stud}, \tilde{y}_{\rm teach}) +  \lambda \cdot {\rm NMR}(S_{\rm stud}, \tilde{S}_{\rm teach}).
\end{equation}
Both terms in the loss function penalise aliasing in the student model output as well as spectral differences between the harmonic distortion components of the student and teacher models. Here we set $\lambda = 1$ with a study on the effect of $\lambda$ left for further work.

\subsection{Training details}
In each training batch the input is a set of sine tones of duration 1.2 seconds. These are processed through the teacher model, truncated to 1 second ($N = F_s$ samples) and anti-aliased to obtain the target batch. For RNN-based models, the first $N/5$ samples were processed through the student model to initialise the states, then truncated back-propagation through time (TBPTT) was implemented with a frame size of $N/10$ samples. For TCN models, the first $N/5$ samples were discarded and the rest processed in frames of $N/2$ samples (due to memory constraints only).
% \SB[OK, these frame sizes seem very large relative to the total signal length (1s), so you are only getting a handful of frames per signal.] 
The batch size was 40 for RNN models and 32 for TCN models. All models were trained using Adam optimizer with a learning rate of 5e-4 for a maximum of 40k iterations (maximum 24 hours on a NVIDIA GeForce GTX 1080). Double precision was used.

\subsection{Validation dataset and metrics}\label{subsec:metrics}
The models were validated and tested using two datasets: an audio signal comprised of 60s of guitar and bass recordings; and a set of sine tones ``playing" the chromatic scale from midi note number 21 to 108 ($f_0$ ranging from of \SI{27.5}{\Hz} to \SI{4186}{\Hz}). Each tone was repeated at three different amplitudes: \SI{-36}{\dB}, \SI{-18}{\dB} and \SI{-6}{\dB}. The signals were passed through the teacher model, student model and reference analog device (if available) to obtain the corresponding $y_{\rm teach}$, $y_{\rm stud}$ and $y_{\rm ref}$. The guitar and bass (G+B) data and the sine tone data are then analysed separately. 

On the G+B data we compute the ESR and a multi-resolution spectral convergence loss (MRSL)
% \SB[Was this defined in the earlier section?]
between log-magnitude mel-spectrograms of the signal and reference \cite{YamatotoMRSL2020, steinmetz2020auraloss}. Where the reference analog device was not available, the teacher model output was used as the reference. For each sine tone input the following metrics are computed on output signal $y$ (either $y_{\rm teach}$ or $y_{\rm stud}$):
\\
\noindent \textbf{ESR-R}: the ESR w.r.t. the reference, $\mathbb{E}(y, y_{\rm ref})$; \\
\noindent \textbf{NMR-R}: the NMR w.r.t. the reference, ${\rm NMR}(S, S_{\rm ref})$;  \\
\noindent \textbf{HESR-R}: the ESR of the magnitude of the harmonic components w.r.t. the reference, $\mathbb{E}(|\tilde{Y}|, |\tilde{Y}_{\rm ref}|)$ where capital $\tilde{Y}$ denotes the bandlimited spectrum; \\
\noindent \textbf{ESR-T}: the ESR w.r.t. the bandlimited teacher as used in the loss function \eqref{eq:loss}, $\mathbb{E}(y, \tilde{y}_{\rm teach})$; \\
\noindent \textbf{NMR-T}: the NMR w.r.t. the bandlimited teacher as used in the loss function \eqref{eq:loss}, ${\rm NMR}(S, \tilde{S}_{\rm teach})$; \\
\noindent \textbf{HESR-T}: the ESR of the magnitude of the harmonic components w.r.t. the teacher, $\mathbb{E}(|{\tilde Y}|, |{\tilde Y}_{\rm teach}|)$; \\
\noindent \textbf{ESR-S}: the ESR of the signal w.r.t. to the bandlimited version of itself, $\mathbb{E}(y, \tilde{y})$; \\
\noindent \textbf{NMR-S}: the NMR of the signal w.r.t. to the bandlimited version of itself, ${\rm NMR}(S, \tilde{S})$. \\
In the calculations above, where the reference device was not available, the bandlimited teacher model output was used as the reference, i.e. $y_{\rm ref} := \tilde{y}_{\rm teach}$. 

\section{Experiments} \label{sec:experiments}
To test our methodology, we implement the teacher-student fine tuning on a total of eight neural models: two open-weight LSTMs, two open-weight TCNs and four models pre-trained by us using data from two different analog distortion effects units (available open-weight at the accompanying webpage).

\subsection{Open-weight models}\label{subsec:guitarml}
The GuitarML Tone Library\footnote{\href{https://guitarml.com/tonelibrary/tonelib-pro.html}{https://guitarml.com/tonelibrary/tonelib-pro.html}} contains open-weight models of various distortion effects, each consisting of an LSTM unit with hidden size 40 followed by a linear layer, with a residual connection between the input and output \cite{Wright2020}. Out of the 18 ``snapshot'' (non-conditioned) models on the homepage, the two ``worst case" models exhibiting the most aliasing (highest NMR-S) were selected as case studies. We refer to these models as \textit{Mesa} and \textit{Goat}.

Neural Amp Modeller (NAM)\footnote{\href{https://www.neuralampmodeler.com/}{https://www.neuralampmodeler.com/}} is an open-source plug-in and framework for training models with hundreds of open-weight models available on Tone3000 \footnote{\href{https://www.tone3000.com/}{https://www.tone3000.com/}}. The two TCN models were selected as those which exhibited the most aliasing out of the ten all-time most downloaded packages (each of which contains several snapshots). The architecture of both is the NAM default: two TCN blocks each with a kernel size of 3, 10 layers, a dilation-growth of 2 and $\tanh$ activation functions. The blocks have channel widths of 16 and 8 respectively. According to the metadata these models were originally trained at $F_s = $ \SI{48}{\kHz}, so fine-tuning was implemented at this same rate. The NAM PyTorch implementation was used with no modifications to the model code. Here we refer to our chosen TCN models as \textit{Vox} and \textit{JCM} with reference to the amplifiers on which they were modelled. 

\subsection{Custom models}
We also trained from scratch models of two analog devices: the Hudson Broadcast germanium pre-amp pedal; and the Dunlop-MXR JHM8 Jimi Hendrix Gypsy Fuzz pedal. For each device we trained both an LSTM model and a TCN model with the architectures described in Sec \ref{subsec:guitarml}. We used the training signal provided by GuitarML, consisting of chirps, noise bursts, and guitar and bass playing amounting to 3 minutes and 40 seconds of audio with $F_s = $~\SI{44.1}{\kHz}. The LSTM training used TBPTT with a warm-up of 1000 samples and a frame-size of 2048 samples. The TCN training used a signal length of 16384 samples. In both cases the batch size was 40 and the models were trained for a maximum of 5k epochs. The loss function was a combination of ESR and DC loss, with A-weighting pre-emphasis filtering \cite{wright2020_perceptual}. 

\begin{table*}[ht!]
\centering
\caption{{\it Mean signal metrics in decibels (lower better) over the audio dataset (col. 4-5) and sine tone dataset (col. 6-15). The sine tone metrics displayed are the arithmetic mean over all input $f_0$ and amplitude. Bolding indicates the best result between the teacher and student models for a given Device and Model. }}

\resizebox{\textwidth}{!}{
\begin{tabular}{ccccccccccccc|cc}
\toprule 
&&& \multicolumn{2}{c}{Audio dataset} & \multicolumn{8}{c}{Sine tone dataset} & \multicolumn{2}{|c}{2x oversampled}\\ 
\cmidrule(lr{0.1em}){4-5}\cmidrule(lr{0.1em}){6-13}\cmidrule(lr{0.0em}){14-15}
Device & Model & Role & G+B ESR & G+B MRSL & ESR-R & NMR-R & HESR-R & ESR-T & NMR-T & HESR-T & ESR-S & NMR-S & ESR-S-OS2 & NMR-S-OS2 \\ 
\midrule 
\multirow{2}{*}{Broadcast} & \multirow{2}{*}{LSTM} & Teacher & {\bf -15.4} & {\bf -18.1} & -11.7 & 8.8 & -14.2 & {\bf -25.5} & 6.1 & $-\infty$ & -25.5 & 6.1 & -44.1 & -3.4 \\ 
 && Student & -12.1 & -12.1 & {\bf -12.6} & {\bf -1.8} & {\bf -15.4} & -19.2 & {\bf -12.6} & -21.3 & {\bf -34.7} & {\bf -16.7} & -52.3 & -23.5 \\ 
\hline 
\multirow{2}{*}{Broadcast} & \multirow{2}{*}{TCN} & Teacher & {\bf -12.1} & {\bf -13.3} & {\bf -12.5} & 7.1 & {\bf -14.6} & {\bf -34.6} & 6.5 & $-\infty$ & -34.6 & 6.5 & -42.3 & -2.1 \\ 
 && Student & -7.0 & -8.3 & -8.2 & {\bf 0.5} & -9.5 & -11.8 & {\bf -1.8} & -13.4 & {\bf -49.3} & {\bf -18.2} & -55.2 & -27.6 \\ 
\hline 
\multirow{2}{*}{JHM8} & \multirow{2}{*}{LSTM} & Teacher & {\bf -24.1} & {\bf -24.7} & {\bf -17.4} & 3.3 & {\bf -19.2} & {\bf -42.7} & -8.9 & $-\infty$ & -42.7 & -8.9 & -57.6 & -24.8 \\ 
 && Student & -20.1 & -20.4 & -16.2 & {\bf 3.1} & -17.0 & -26.2 & {\bf -16.6} & -27.8 & {\bf -46.8} & {\bf -22.7} & -57.9 & -28.9 \\ 
\hline 
\multirow{2}{*}{JHM8} & \multirow{2}{*}{TCN} & Teacher & {\bf -19.9} & {\bf -21.8} & {\bf -12.0} & 3.9 & -16.1 & {\bf -42.1} & -1.0 & $-\infty$ & -42.1 & -1.0 & -54.9 & -11.8 \\ 
 && Student & -13.5 & -14.1 & -11.7 & {\bf -1.3} & {\bf -16.7} & -22.8 & {\bf -8.5} & -24.5 & {\bf -52.6} & {\bf -25.1} & -58.9 & -29.1 \\ 
\hline 
\multirow{2}{*}{Goat} & \multirow{2}{*}{LSTM} & Teacher & $-\infty$ & $-\infty$ & $-\infty$ & {\bf -122.0} & $-\infty$ & {\bf -29.9} & 11.5 & $-\infty$ & -29.9 & 11.5 & -35.6 & 0.4 \\ 
 && Student & -11.1 & -9.3 & -11.9 & -7.7 & -12.6 & -12.0 & {\bf -6.7} & -12.6 & {\bf -40.8} & {\bf -5.6} & -41.6 & -5.6 \\ 
\hline 
\multirow{2}{*}{Mesa} & \multirow{2}{*}{LSTM} & Teacher & $-\infty$ & $-\infty$ & $-\infty$ & {\bf -126.2} & $-\infty$ & {\bf -29.4} & 19.3 & $-\infty$ & -29.4 & 19.3 & -46.7 & 4.2 \\ 
 && Student & -12.7 & -11.3 & -17.3 & -8.0 & -20.8 & -17.7 & {\bf -10.3} & -20.8 & {\bf -36.4} & {\bf -14.8} & -58.2 & -25.3 \\ 
\hline 
\multirow{2}{*}{Vox} & \multirow{2}{*}{TCN} & Teacher & $-\infty$ & $-\infty$ & $-\infty$ & {\bf -125.4} & $-\infty$ & {\bf -36.8} & 15.0 & $-\infty$ & -36.8 & 15.0 & -45.2 & 6.5 \\ 
 && Student & -8.1 & -10.4 & -17.6 & -4.1 & -20.2 & -17.8 & {\bf -9.0} & -20.2 & {\bf -53.5} & {\bf -21.1} & -56.4 & -24.3 \\ 
\hline 
\multirow{2}{*}{JCM} & \multirow{2}{*}{TCN} & Teacher & $-\infty$ & $-\infty$ & $-\infty$ & {\bf -124.3} & $-\infty$ & {\bf -33.6} & 13.4 & $-\infty$ & -33.6 & 13.4 & -46.1 & 3.0 \\ 
 && Student & -10.0 & -13.2 & -18.1 & -5.2 & -22.6 & -18.2 & {\bf -9.4} & -22.6 & {\bf -44.9} & {\bf -12.5} & -58.7 & -24.6 \\ 
\hline 

\bottomrule 
\end{tabular}
}
\label{tab:resuls_big}
\end{table*}
\section{Objective results} \label{sec:obj_results}
This section presents the objective results along with analysis of example spectra of model outputs. The metrics described in Sec. \ref{subsec:metrics} are reported in Table \ref{tab:resuls_big} for all the models considered.  

\subsection{Custom pre-training results}
The metrics measured on our custom pre-trained models are shown in rows 1, 3, 5 and 7 of Table \ref{tab:resuls_big} (Broadcast/JHM8 Teacher). In terms of G+B ESR and G+B MRSL, the LSTM models gave the better result over the TCN models. Between the two devices, the JHM8 proved the easier device to model, with the JHM8 LSTM giving overall the best result (an ESR of 0.3\%). The sine tone dataset metrics with respect to the reference (-R) are generally higher (worse), which is perhaps unsurprising as there were no pure sine tones in the training data (but there were sine sweeps). In all four models, the NMR-R is greater than zero, suggesting the model outputs are significantly noisier than outputs from the reference device. The NMR-R, however, will also pick up differences between non-noise components. 

\subsection{Fine-tuning results -- aliasing reduction}
The spectral response to a sinusoidal input can be seen in Fig. \ref{fig:broadcast_gypsy_spec} for the Broadcast and JHM8 models and Fig. \ref{fig:multi_spec} for the open-weight models. In all cases, a reduction in aliasing can clearly be observed between the Teacher and Student models. The reduction in aliasing is especially visible in lower frequencies. Aliases below the fundamental are most likely to be audible \cite{Lehtonen_2012_saw}, so it is reassuring that these appear to be the most suppressed. In Fig. \ref{fig:broadcast_gypsy_spec}a, for example, the most prominent sub-fundamental aliases have been reduced in magnitude by approximately \SI{40}{\dB}. This suppression can also be seen in Fig. \ref{fig:broadcast_gypsy_spec}b and Fig. \ref{fig:multi_spec}, with the most extreme example shown in Fig. \ref{fig:multi_spec}a for the \textit{Goat} LSTM model.

The NMR-S results in Table \ref{tab:resuls_big} provide a more objective (yet perceptually informed) metric of aliasing. For all eight models, the fine-tuning process results in a decrease in mean NMR-S across the sinusoids dataset. Because this is an arithmetic mean across all frequencies and input gains, it is useful to examine how NMR-S varies with input frequency -- as shown in Fig. \ref{fig:metrics_vs_f0}a-i for the Broadcast models and \ref{fig:metrics_vs_f0}b-i for the JHM8 models. In Fig. \ref{fig:metrics_vs_f0}a-i, for example, the NMR of the LSTM and TCN Teacher models both follow a similar trajectory; with results exceeding \SI{-10}{\dB} (an approximate threshold of aliasing audibility \cite{Lehtonen_2012_saw, Kahles2019}) for $f_0 \gtrapprox $~\SI{1}{\kHz}. The proposed fine tuning method results in the NMR-R of the Student models being reduced to below \SI{-10}{\dB} for all $f_0$.

\subsection{Fine tuning results -- harmonic analysis}
While it is clear that the proposed method reduces aliasing, it is important to analyse how the desired harmonic distortion components are affected by the fine-tuning process. Ideally, the process would remove all aliasing whilst retaining the exact same amplitudes of harmonics. In practice, there is inevitably some side-effect on the harmonics. The proposed loss function \eqref{eq:loss} was chosen so that it not only penalises aliasing but changes in harmonic distortion components with respect to the Teacher model. The ESR-T and NMR-T are those used in the loss function, so it is interesting to observe how these vary between the Teacher and Student models in Table \ref{tab:resuls_big}. In all cases, the fine-tuning process results in an increase in ESR-T and a decrease in NMR-T. Ideally, these should both decrease, but it appears that the models are trading off between these two measures in training. One would hope that this  means that the Student model is learning to ``shift'' the aliasing/noise from being perceptible (measured by the NMR) to less perceptible (measured by the ESR). However, as shown in the HESR-R results, there is always some change in the amplitudes of the harmonic components.

% \begin{figure}[ht!!]
% \center
% \includegraphics[width=0.5\textwidth]{figs/broadcast_spectra_f0=2349.3.pdf}
% \caption{\label{fig:broadcast_spec}{\it Output magnitude spectra of the Broadcast Teacher models (a-b), their respective Student models (c-d) and the reference (e) for an input tone of \SI{2394.3}{\Hz}.  Crosses mark the harmonic components, and (f) shows the error in magnitude of these w.r.t. the reference.}}
% \end{figure}
\begin{figure*}[ht!!!!]

\begin{minipage}[b]{.5\linewidth}
  \centering
  \centerline{\includegraphics[width=\linewidth, trim={0, 0.3cm, 0, 0.2cm}, clip]{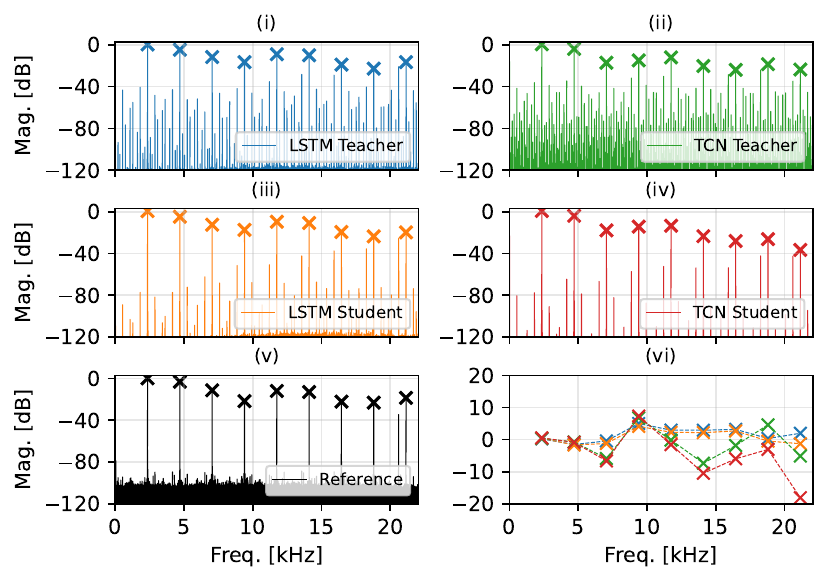}}
%  \vspace{1.5cm}
  \centerline{\footnotesize{{\it (a) Broadcast}}}
\end{minipage}
\hfill
\begin{minipage}[b]{0.5\linewidth}
  \centering
  \centerline{\includegraphics[width=\linewidth, trim={0, 0.3cm, 0, 0.2cm}, clip]{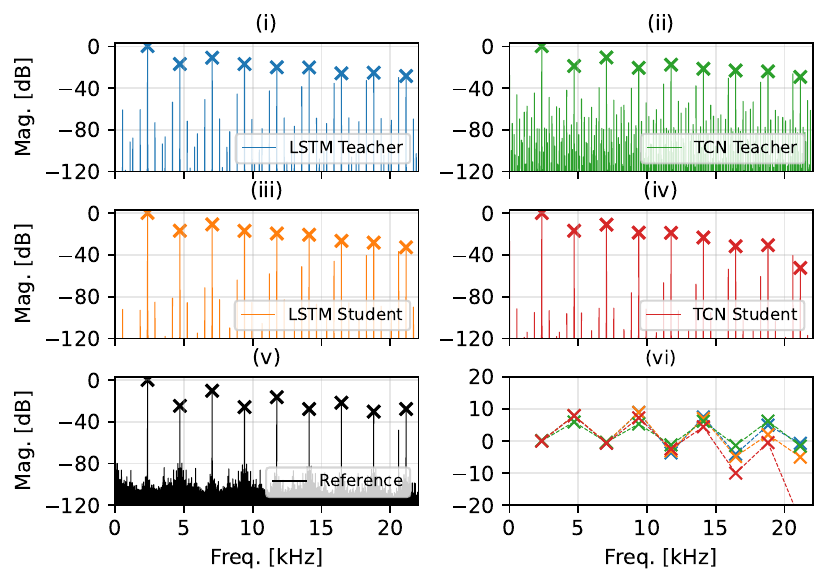}}
%  \vspace{1.5cm}
  \centerline{{\it \footnotesize{(b) JHM8}}}
\end{minipage}
\caption{{\it Output magnitude spectra of the Broadcast (a) and JHM8 (b) Teacher models (i-ii), their respective Student models (iii-iv) and the reference (v) for an input tone of \SI{2394.3}{\Hz}. Crosses mark the harmonic components, and (vi) shows the error in magnitude of these w.r.t. the reference.}}
\label{fig:broadcast_gypsy_spec} 
\end{figure*}
\begin{figure*}[h]
\center
\includegraphics[width=\textwidth, clip, trim={0mm, 2mm, 0mm, 2mm}]{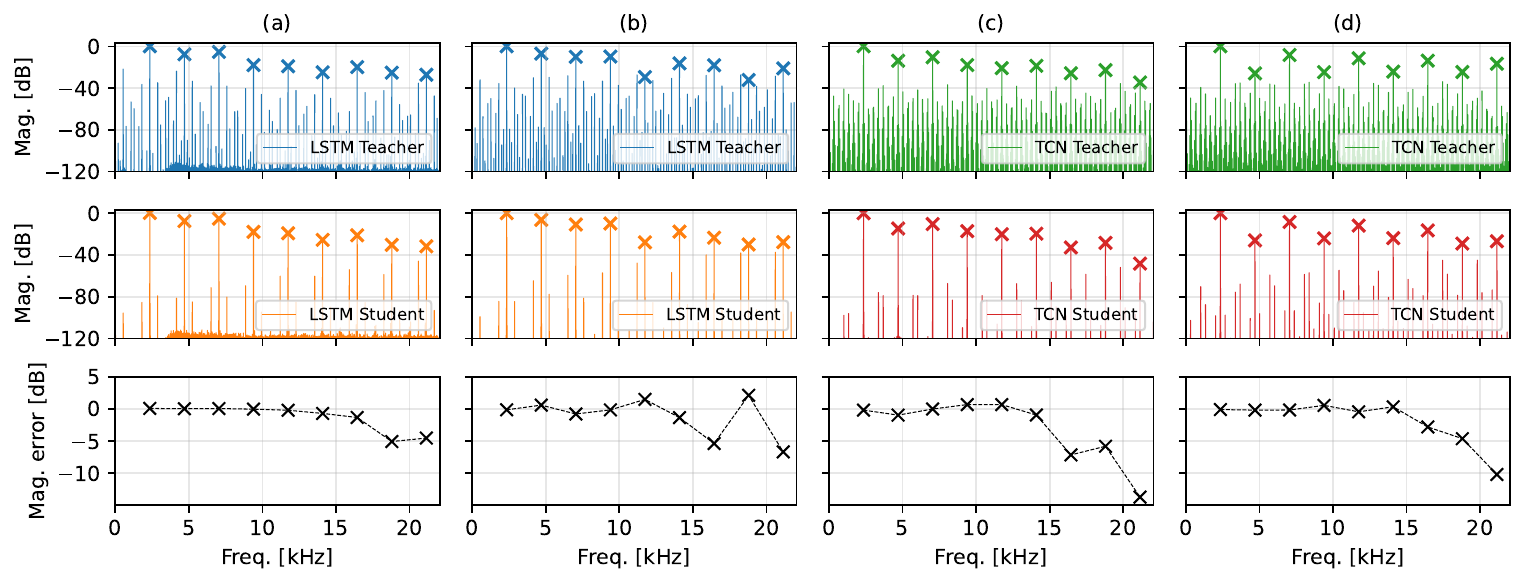}
\caption{\label{fig:multi_spec}{\it Output magnitude spectra of the Teacher models (top), the corresponding Student models (middle) and the relative error in magnitude of the harmonic components (bottom) for the open-weight Goat (a), Mesa (b), Vox (c) and JCM (d) models. The input tone had $f_0$ = \SI{2394.3}{\Hz} and amplitude \SI{-6}{\dB}.}}
\end{figure*}

The effect on the harmonic components for a sinusoidal input can be seen in Figures \ref{fig:broadcast_gypsy_spec} and \ref{fig:multi_spec}. The general trend -- especially noticeable in Fig. \ref{fig:multi_spec} -- is that the anti-aliasing procedure results in a damping of high frequency harmonics. For frequency components above \SI{12}{\kHz}, this is to be fully expected due to the pre-emphasis LPF used during training -- errors in high frequency harmonics are ignored by design. Errors in harmonics within the more sensitive human hearing bands are more critical as they may be perceived as a difference in the desired harmonic distortion.

\begin{figure*}[ht]

\begin{minipage}[b]{.5\linewidth}
  \centering
  \centerline{\includegraphics[width=\linewidth, trim={0, 0.3cm, 0, 0.2cm}, clip]{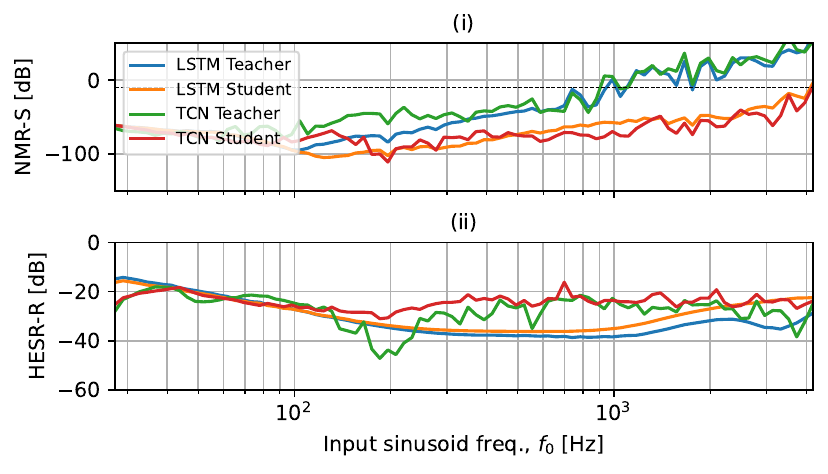}}
%  \vspace{1.5cm}
  \centerline{\footnotesize{{\it (a) Broadcast}}}
\end{minipage}
\hfill
\begin{minipage}[b]{0.5\linewidth}
  \centering
  \centerline{\includegraphics[width=\linewidth, trim={0, 0.3cm, 0, 0.2cm}, clip]{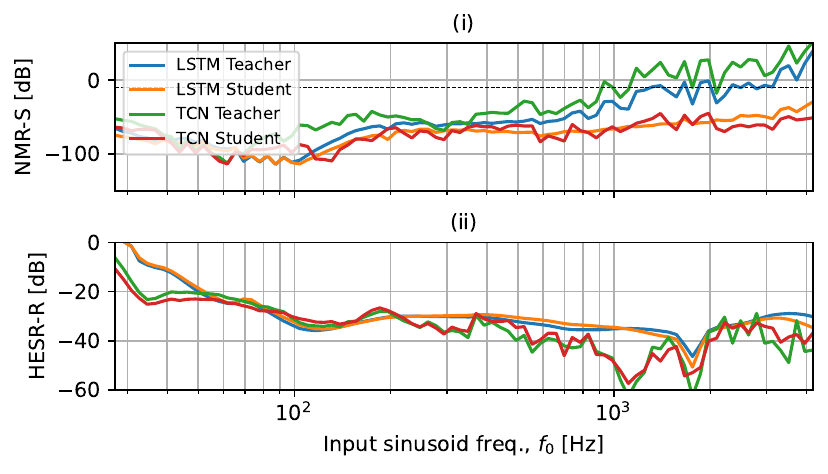}}
%  \vspace{1.5cm}
  \centerline{{\it \footnotesize{(b) JHM8}}}
\end{minipage}
\caption{{\it NMR-S (i) and HESR-R (ii) against input sinusoidal frequency, $f_0$, for the Broadcast (a) and JHM8 (b) models. The input gain was \SI{-6}{\dB}. The black dashed line at \SI{-10}{\dB} indicates the approximate threshold of aliasing audibility \cite{Lehtonen_2012_saw}. Lower values are better.}}
\label{fig:metrics_vs_f0} 
\end{figure*}

For the Broadcast and JHM8 models, the magnitude error in harmonic components with respect to the reference (the HESR-R) is shown in Table \ref{tab:resuls_big}. Interestingly, in some cases (Broadcast LSTM and JHM8 TCN) the Student models achieve a lower (better) HESR-R than their respective Teacher models, which is a remarkable result considering the Student models were shown no additional data from the reference device during fine-tuning. In the other two cases, however, the result is the opposite and the largest discrepancy can be seen between the Broadcast TCN models. Fig. \ref{fig:metrics_vs_f0}a-ii shows HESR-R against input sinusoidal frequency for the Broadcast models. The results may appear similar at first glance, but there is a large discrepancy between the TCN Teacher and Student models for $f_0$ around \SI{200}{\Hz} -- a perceptually important frequency range for guitar and bass processing.

\subsection{Comparison with oversampling}
It is interesting to compare the results of the fine-tuning process with that of oversampling the original pre-trained model. Here we use an oversampling factor of two, and implement this for both the Teacher and Student models in all cases. The oversampled LSTMs were implemented via the method in \cite{Carson2024}. For the TCN models, oversampling was implemented by upsampling the convolutional kernels, i.e. increasing the convolution dilation of each layer by a factor of $2$. In all cases, frequency-domain resampling was used to convert the sample rate of the input/output signals to/from the oversampled rate (see e.g. \cite{Valimaki2023}).

\begin{figure}[ht!]
\center
\includegraphics[width=0.5\textwidth, clip, trim={0mm, 2.5mm, 0mm, 2mm}]{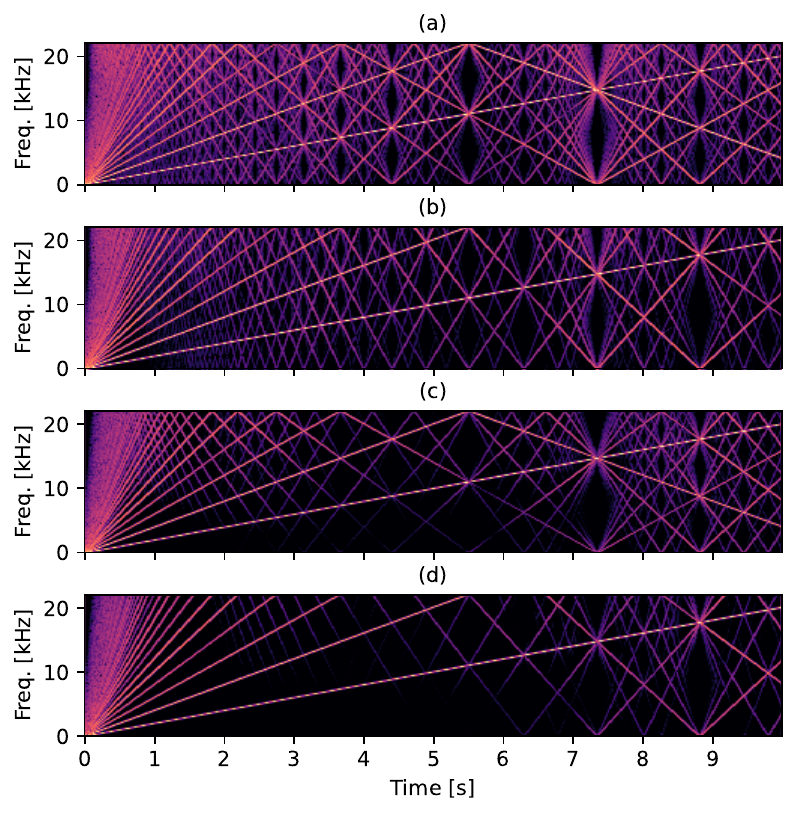}
\caption{\label{fig:broadcast_sweep}{\it Response to a sine sweep for the Broadcast LSTM models: (a) Teacher (b) Teacher 2x oversampled (c) Student (d) Student 2x oversampled. The minimum amplitude visible is \SI{-80}{\dB}.}}
\end{figure}

The ESR-S and NMR-S for the oversampled models are shown in the last two columns of Table \ref{tab:resuls_big}. In all but one case (JHM8 LSTM) the proposed fine-tuned model performs better in terms of aliasing reduction than 2x oversampling of the original model. The fine-tuning process requires extra resources during training, but at inference it requires no extra operations per sample compared to the original model, unlike oversampling. Fig. \ref{fig:broadcast_sweep} shows a sine sweep passed through the Broadcast LSTM Teacher and Student models both at the base and oversampled rates.

\section{Perceptual Evaluation}\label{sec:mushra}
A Multiple Stimuli Hidden Reference and Anchor (MUSHRA) \cite{MUSHRA_ITU-R} listening test was conducted to investigate how the original pre-trained models (Teacher) and the fine-tuned (Student) models were perceived compared to the reference analog audio device. Only the \textit{Broadcast} and \textit{JHM8} models were included in the test; the others were excluded due to the lack of reference audio data. The anchor was the input signal hard-clipped between -0.5 and 1.0 with an input and output gain of \SI{48}{\dB} and\SI{-9}{\dB} respectively. 

Six input signals were used to generate the test excerpts: two direct-input (DI) electric guitar clips, two DI bass clips, a linear sine sweep from \SI{20}{\Hz} to \SI{10}{\kHz} and a linear sine sweep from \SI{10}{\kHz} to \SI{20}{\kHz}. Across the two reference devices, this gave 12 trials per test. For each trial, listeners were presented with an audio clip from the reference device and asked to rate the four model outputs (LSTM Teacher, LSTM Student, TCN Teacher and TCN Student), the Hidden Reference and the Anchor based on their perceived similarity to the reference.

Sixteen volunteers participated in the test, all of which were either students, academics, professionals or practitioners within audio technology. Participants who rated the Reference < 80\% in > 15\% of trials were excluded from the analysis, leaving 12 remaining. This is a relaxation of the  MUSHRA standard post-screening guidance \cite{MUSHRA_ITU-R} in which a similarity threshold of 90\% is recommended, but using this criterion there would have only been six participants remaining. 

\begin{figure}[ht!]
    \centering
    \includegraphics[width=\linewidth, clip, trim={0mm, 2.5mm, 0mm, 2mm}]{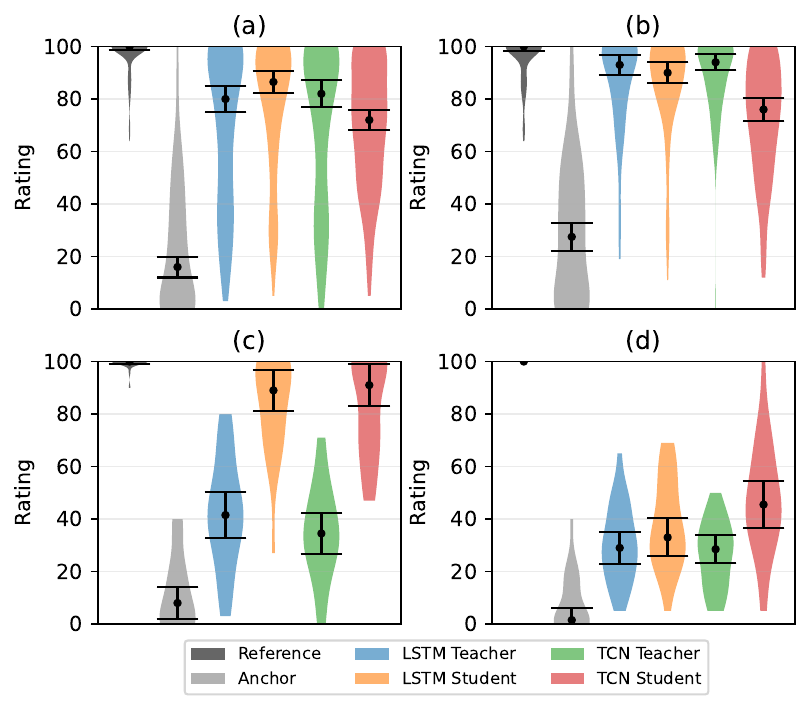}
    \caption{{\it Perceived similarity ratings across (a) all samples combined, (b) guitar and bass samples, (c) the sine sweep from 0 to 10kHz and (d) the sine sweep from 10k-20kHz. Results from the Broadcast and JHM8 pedals are aggregated together. Black dots and error bars show the median and 95\% confidence interval.}}
    \label{fig:mushra}
\end{figure}

Violin plots of the MUSHRA test results aggregated across participants and both devices are shown in Fig. \ref{fig:mushra}. The medians with 95\% confidence intervals (CI) are displayed. 

While the participants were generally capable of identifying the Reference, there were some cases where they rated it less than 100 -- suggesting confusion over which was the correct Reference. These cases were much more common for guitar and bass inputs than for the sweeps. The Anchor results show a large spread, and again this was input dependent. For the sweep inputs, the Anchor was exclusively rated lower than 40 with the median and 95\% confidence interval below 20 (``Very Poor'' perceived similarity). For guitar and bass inputs, the results were higher and in some cases the anchor was rated very highly -- suggesting perhaps this was not the best choice of anchor for the experiment.

The results for the original Teacher models (both LSTM and TCN) show a large spread of results, and a bimodal distribution is observed in Fig. \ref{fig:mushra}a. For guitar and bass inputs, the median CIs for both the LSTM Teacher and TCN model lie above 85\% -- indicating that both models give an Excellent perceived similarity to the reference. Since the reference contains no aliasing, it appears that aliasing caused by the original models was not noticeable for the guitar and bass inputs. 
When driven with the sine sweeps however, the perceived similarity is significantly lower with the median CI within the Very Poor/Fair rating bands. Participants reported that under these conditions, it was much easier to distinguish models from the reference due to the presence of aliasing artefacts. 

The results for the fine-tuned (Student) models also show a distinction between the input stimuli. For the sine sweep up to 10k, there is a significant increase in perceived similarity compared to the Teacher models, with the median and CI in the Excellent range for both LSTM and TCN Student models. This shows that the proposed models have been effective at reducing perceived aliasing for inputs within this bandwidth. For frequencies above \SI{10}{\kHz}, however, there is less improvement. The TCN Student model shows a statistically significant improvement over its Teacher model but the median CI is only within the Fair band, whereas for the LSTM there is very little improvement. Considering the models were trained on sinusoids up to a maximum $f_0$ of \SI{12}{\kHz}, it is not surprising that aliasing is still noticeable for input frequencies beyond this. On the guitar and bass stimuli, the LSTM Student model shows no significant change in perceived quality compared to its Teacher. This is a good result: where the original model performed well, its performance has not been affected by fine-tuning; but where aliasing was originally a problem, the fine-tuning process has reduced aliasing (at least for inputs up to \SI{10}{\kHz}). The TCN Student model results, however, show a significant reduction in perceived similarity to the Reference for guitar and bass inputs. It was found that the measured suppression in high frequencies (Sec. \ref{sec:obj_results}) were indeed perceptible, as some participants who reported that the TCN Student models sounded ``low-passed'' or ``less distorted'' than the reference. However, the median TCN Student score and CI was still within the Very Good range.

\section{Conclusions and further work}\label{sec:conclusion}
This work presented a fine-tuning procedure for reducing the aliasing caused by neural network models of guitar distortion effects. This involved training a Student model against an aliasing-free synthetic dataset --generated during training by processing sinusoids through the original pre-trained (Teacher) model and then removing non-harmonic components through Fourier analysis and re-synthesis. As case studies, open-weight LSTM and TCN models were considered, and an example of each trained from scratch on two analog fuzz effects pedals. The proposed method consistently reduced aliasing across all systems, outperforming two times oversampling in all but one case. However, fine-tuning sometimes altered desirable harmonic content.  A MUSHRA listening test was deployed to evaluate how the original pre-trained (Teacher) models and the fine-tuned (Student) models compared in perceived similarity to two analog reference devices. It was found that for sine sweep inputs -- for which lots of aliasing was present in the Teacher outputs -- fine-tuning significantly improved the similarity score for both LSTM and TCN models. For non-sinusoidal guitar and bass signals, there was no significant difference between the LSTM Student model and its Teacher, with both rated as Excellent in similarity to the reference. For the TCN models, there was a reduction in perceived similarity from Excellent to Very Good, indicating that fine-tuning had an adverse effect on the desired harmonic distortion. While our results show the potential of the proposed method for anti-aliasing in neural networks, there are still areas for further work, particularly regarding the affected harmonic distortion components. For example a hyper-parameter sweep of loss function weighting $\lambda$, a comparison of different pre-emphasis filters or an investigation into corrective filters post-training.

\bibliographystyle{IEEEbib}
\footnotesize
\bibliography{DAFx25_tmpl} % requires file DAFx25_tmpl.bib
\end{document}